\documentstyle[prd,preprint,tighten,aps,eqsecnum,amssymb,epsfig]{revtex}
\begin{document}
\draft
\preprint{
\begin{tabular}{r}
DFTT 25/95
\\
JHU-TIPAC 95013
\\
hep-ph/9504405
\end{tabular}
}
\title{NEUTRINO OSCILLATIONS
IN THE FRAMEWORK OF THREE-GENERATION MIXINGS
WITH MASS HIERARCHY}
\author{
S.M. Bilenky$^{\mathrm{a,b}}$,
A. Bottino$^{\mathrm{a}}$,
C. Giunti$^{\mathrm{a}}$
and
C. W. Kim$^{\mathrm{c}}$
}
\address{
\begin{tabular}{c}
$^{\mathrm{a}}$INFN,
Sezione di Torino and Dipartimento di Fisica Teorica,
Universit\`a di Torino,
\\
Via P. Giuria 1, 10125 Torino, Italy.
\\
$^{\mathrm{b}}$Joint Institute of Nuclear Research,
Dubna, Russia.
\\
$^{\mathrm{c}}$Department of Physics and Astronomy,
The Johns Hopkins University,
\\
Baltimore, Maryland 21218, USA.
\end{tabular}
}
\date{\today}
\maketitle
\begin{abstract}
We have analyzed the results of
reactor and accelerator
neutrino oscillation experiments
in the framework of a general model
with mixing of three neutrino fields
and
a neutrino mass hierarchy that
can accommodate the results of
the solar neutrino experiments.
It is shown that
$ \nu_\mu \leftrightarrows \nu_e $
oscillations
with
$ 0.6 \le \Delta m^2 \le 100 \, \mathrm{eV}^2 $
and amplitude larger than
$ 2 \times 10^{-3} $
are not compatible with the existing limits
on neutrino oscillations
if the non-diagonal elements
of the mixing matrix
$ \left| U_{e3} \right| $
and
$ \left| U_{\mu3} \right| $
are small.
Thus,
if the excess of electron events
recently observed in the LSND experiment
is due to
$ \nu_\mu \leftrightarrows \nu_e $
oscillations,
the mixing
in the lepton sector
is basically different from the CKM mixing
of quarks.
If this type of mixing is realized in nature,
the observation of
$ \nu_\mu \leftrightarrows \nu_e $
oscillations
would not influence
$ \nu_\mu \leftrightarrows \nu_\tau $
oscillations
that are being searched for in the CHORUS and
NOMAD experiments.
\end{abstract}

\pacs{}

\narrowtext

\section{Introduction}
\label{INTRO}

The problem of neutrino mass and mixing
is one of the major issues that confront us
in neutrino physics at present.
The search for the effects of neutrino mass
is generally considered as
one of the promising ways to probe
new physics beyond the standard model.
Many experiments
are currently under way in different laboratories
in order to investigate the effects of neutrino mixing
and the nature
of massive neutrinos
(Dirac or Majorana?).

Important indications
in favor of neutrino masses and mixing
have been obtained in the solar neutrino experiments.
The event rates measured
in the all four solar neutrino experiments
\cite{SOLAREXP},
which are sensitive to different parts
of the solar neutrino spectrum,
are significantly lower
than the event rates predicted by the
Standard Solar Model
(see Ref.\cite{BAHCALL}).
Moreover,
a phenomenological analysis
(see Ref.\cite{PHENOMEN})
that does not depend on the predictions
of the Standard Solar Model
shows that the data from different experiments
cannot be accommodated simultaneously
if we assume that solar $\nu_e$'s
are not transformed into other states.

Another indication in favor
of non-zero neutrino mass
comes from cosmology.
The most plausible cosmological scenario
of dark matter,
which can describe the COBE data
and large scale structures  of the universe,
is a mixture of cold dark matter
with 20--30\% of hot dark matter
(see Ref.\cite{DARK}).
In this scenario
it is conjectured that the neutrinos
which form the hot dark matter
have mass
of the order of several eV.

New neutrino oscillation experiments
searching for the effects of neutrino mass
in the cosmological range
are under way.
The
CHORUS
\cite{CHORUS}
and NOMAD
\cite{NOMAD}
experiments are looking for
$ \nu_\mu \to \nu_\tau $
oscillations
and the
KARMEN
\cite{KARMEN}
and
LSND
\cite{LSND}
experiments
are searching for
$ \nu_\mu \to \nu_e $
oscillations.
It was reported recently
\cite{LSND}
that
an excess of 9 electron events
with an expected background of
$ 2.1 \pm 0.3 $
events was observed in the LSND experiment.
If these events are due to
$ \bar\nu_\mu \to \bar\nu_e $
oscillations,
the average
$ \bar\nu_\mu \to \bar\nu_e $
transition probability
is equal to
$ 0.34^{+0.20}_{-0.18}\pm0.07\% $.

We analyze here
the results of the experiments that have searched for
neutrino oscillations
with reactor and accelerator neutrinos
in the framework of a general model
with mixing of three neutrino fields.
We only assume that the neutrino masses
satisfy a hierarchy that can accommodate
the solar neutrino data.
We show that a positive signal
in the LSND experiment
would mean that
in the lepton sector
there is no usual hierarchy
of couplings among generations.
We also discuss
implications of a possible positive signal
in the LSND experiment on
$ \nu_\mu \to \nu_\tau $
oscillation experiments.

\section{Three-generations mixing with neutrino mass hierarchy}
\label{SCHEME}

We will discuss
oscillations of terrestrial neutrinos
in the framework of a scheme of mixing
of three massive neutrino fields
assuming only a hierarchy of neutrino masses.

According to the general theory of neutrino mixing
(see, for example, Refs.\cite{BILENKY,CWKIM}),
the left-handed flavor neutrino fields
$\nu_{\alpha L}$
are superpositions
of the left-handed components
of (Dirac or Majorana)
massive neutrino fields
$\nu_{kL}$:
\begin{equation}
\nu_{\alpha L}
=
\sum_{k=1}^{n}
U_{\alpha k}
\nu_{kL}
\;.
\label{E100}
\end{equation}
Here $U$ is a unitary mixing matrix.
{}From LEP data
(see Ref.\cite{RPP})
it follows that
three neutrino flavors exist in nature.
The number $n$ of massive neutrinos
depends on the scheme of neutrino mixing.
In the case of a Dirac mass term,
the total lepton number is conserved,
massive neutrinos are Dirac particles
and
$n=3$.
In the case of a Majorana mass term,
only left-handed flavor neutrino fields
enter in the Lagrangian,
massive neutrinos are Majorana particles
and
$n=3$.
In the most general case
of a Dirac and Majorana mass term,
massive neutrinos are Majorana particles
and
$n=6$.
In this last case
there is a very attractive possibility
for the generation of neutrino masses,
i.e.
the see-saw mechanism
\cite{SEESAW}.
This mechanism is based on the assumption
that the conservation of the total lepton number
is violated at some large energy scale $M$.
In this case,
in the spectrum of Majorana masses
there are three light masses
$m_k$
and three heavy masses,
of the order of $M$.
In the simplest see-saw model,
the neutrino masses
are given by
$
m_k
\simeq
m_{\mathrm{F}k}^2
/
M
$
($k=1,2,3$),
where
$ m_{\mathrm{F}k} $
is the mass of the up-quark or charged lepton
in the $k^{\mathrm{th}}$ generation.
The see-saw mechanism
naturally explains the experimental fact
that the neutrino masses are much smaller than those
of other fundamental fermions.

We will consider here
a scheme of mixings among three massive neutrino fields
assuming that the neutrino masses satisfy
the hierarchy relation
\begin{equation}
m_1 \ll m_2 \ll m_3
\label{E103}
\end{equation}
which is suggested by
the see-saw mechanism.

In the following
we will not impose any theoretical constraints
on the elements of the mixing matrix $U$.
{}From Eq.(\ref{E100}) it follows that
the state
$
\left|
\nu_{\alpha}
\right\rangle
$
of a flavor neutrino with momentum $p$
is given by
\begin{equation}
\left|
\nu_{\alpha}
\right\rangle
=
\sum_{k=1}^{3}
U_{\alpha k}^{*}
\left|
\nu_{k}
\right\rangle
\;,
\label{E104}
\end{equation}
where
$
\left|
\nu_{k}
\right\rangle
$
is the state of a neutrino
with definite mass $m_k$.
It is to be stressed that the state of a neutrino
with a definite flavor
is not a state with definite mass and
the notion of
mass of a flavor neutrino
can have only an approximate meaning.

The amplitude of
$ \nu_{\alpha} \to \nu_{\beta} $
transitions
can be written in the following form:
\begin{equation}
\cal{A}_{\nu_{\alpha}\to\nu_{\beta}}
=
{\mathrm{e}}^{ - i E_1 L }
\left\{
\sum_{k=2}^{3}
U_{\beta k}
\left[
\exp
\left(
- i
{\displaystyle
\Delta m^2_{k1} L
\over\displaystyle
2 p
}
\right)
- 1
\right]
U_{\alpha k}^{*}
+
\delta_{\beta\alpha}
\right\}
\;.
\label{E201}
\end{equation}
Here
$L$ is the distance between
the neutrino source and detector,
$ \Delta m^2_{k1} \equiv m^2_k - m^2_1 $,
$p$ is the neutrino momentum
and
$ \displaystyle
E_k
=
\sqrt{ p^2 + m_k^2 }
\simeq
p
+
{\displaystyle
m_k^2
\over\displaystyle
2 p
}
$.

We will assume that
$
\Delta m^2_{21}
\ll
\Delta m^2_{31}
$
is relevant for solar neutrinos,
say
$ \Delta m^2_{21} \simeq 5 \times 10^{-6} \, \mathrm{eV}^2 $,
as suggested by the MSW interpretation of the solar
neutrino deficit.
Thus,
for experiments with terrestrial neutrinos
$ \Delta m^2_{21} L / 2 p \ll 1 $
and
the probability of
$ \nu_{\alpha} \to \nu_{\beta} $
($ \bar\nu_{\alpha} \to \bar\nu_{\beta} $)
transitions with $\beta\not=\alpha$
is given by
(see, for example, Ref.\cite{BFP92})
\begin{equation}
P_{\nu_{\alpha}\to\nu_{\beta}}
=
P_{\bar\nu_{\alpha}\to\bar\nu_{\beta}}
=
{1\over2}
\,
A_{\nu_{\alpha};\nu_{\beta}}
\left(
1
-
\cos
{\displaystyle
\Delta m^2_{31} \, L
\over\displaystyle
2 \, p
}
\right)
\; ,
\label{E107}
\end{equation}
where
\begin{equation}
A_{\nu_{\alpha};\nu_{\beta}}
=
A_{\nu_{\beta};\nu_{\alpha}}
=
4
\left| U_{\alpha3} \right|^2
\left| U_{\beta3} \right|^2
\label{E108}
\end{equation}
is the
amplitude of
$ \nu_{\alpha} \leftrightarrows \nu_{\beta} $
($ \bar\nu_{\alpha} \leftrightarrows \bar\nu_{\beta} $)
oscillations.
Let us stress that Eq.(\ref{E107})
has the same form as the expression
for the oscillation probability
in the case of mixing between two generations.
In this last case,
the oscillation amplitude is equal to
$ \sin^2 2 \vartheta $,
where $ \vartheta $ is the mixing angle.

The survival probability of $\nu_\alpha$ ($\bar\nu_\alpha$)
can be obtained from the unitarity constraint to be
\begin{equation}
P_{\nu_{\alpha}\to\nu_{\alpha}}
=
P_{\bar\nu_{\alpha}\to\bar\nu_{\alpha}}
=
1
-
\sum_{\beta\not=\alpha}
P_{\nu_{\alpha}\to\nu_{\beta}}
=
1
-
{1\over2}
\,
B_{\nu_{\alpha};\nu_{\alpha}}
\left(
1
-
\cos
{\displaystyle
\Delta m^2_{31} \, L
\over\displaystyle
2 \, p
}
\right)
\;,
\label{E109}
\end{equation}
where
the oscillation amplitude
$ B_{\nu_{\alpha};\nu_{\alpha}} $
is given by
\begin{equation}
B_{\nu_{\alpha};\nu_{\alpha}}
=
\sum_{\beta\not=\alpha}
A_{\nu_{\alpha};\nu_{\beta}}
\;.
\label{E112}
\end{equation}
Using the unitarity of the mixing matrix we obtain
\begin{equation}
B_{\nu_{\alpha};\nu_{\alpha}}
=
4
\left| U_{\alpha3} \right|^2
\left(
1
-
\left| U_{\alpha3} \right|^2
\right)
\label{E110}
\end{equation}
We wish to emphasize the following important features of
terrestrial neutrino oscillations
in the model under consideration:

\begin{enumerate}

\item
All channels
($\nu_{\mu}\leftrightarrows\nu_{e}$,
 $\nu_{\mu}\leftrightarrows\nu_{\tau}$,
 $\nu_{e}\leftrightarrows\nu_{\tau}$)
are open if all the elements
$ U_{\alpha3} $
are different from zero.

\item
The oscillations in all channels
are characterized
by the {\em same} oscillation length
$
L_{\mathrm{osc}}
=
4 \pi p / \Delta m^2_{31}
$.

\item
The equalities
$
P_{\nu_{\alpha}\to\nu_{\beta}}
=
P_{\bar\nu_{\alpha}\to\bar\nu_{\beta}}
$
for $\alpha\not=\beta$
are satisfied even if CP is not conserved
in the lepton sector.

\item
The oscillation amplitudes
$ A_{\nu_{\alpha};\nu_{\beta}} $
satisfy the following relation:
\begin{equation}
\left(
{\displaystyle
A_{\nu_{\mu};\nu_{e}}
A_{\nu_{\mu};\nu_{\tau}}
\over\displaystyle
A_{\nu_{e};\nu_{\tau}}
}
\right)^{1/2}
+
\left(
{\displaystyle
A_{\nu_{\mu};\nu_{e}}
A_{\nu_{e};\nu_{\tau}}
\over\displaystyle
A_{\nu_{\mu};\nu_{\tau}}
}
\right)^{1/2}
+
\left(
{\displaystyle
A_{\nu_{e};\nu_{\tau}}
A_{\nu_{\mu};\nu_{\tau}}
\over\displaystyle
A_{\nu_{\mu};\nu_{e}}
}
\right)^{1/2}
=
2
\;.
\label{E202}
\end{equation}
This relation follows from Eq.(\ref{E108})
and the unitarity of the mixing matrix.

\end{enumerate}

\section{Discussion
of the results of neutrino oscillation experiments}

In this section
we will consider the results
of the experiments that have searched for
oscillations of terrestrial neutrinos
in the framework of the scheme
presented in Section \ref{SCHEME}.

{}From the unitarity of the mixing matrix it follows that
$
\left| U_{\tau3} \right|^2
=
1
-
\left| U_{e3} \right|^2
-
\left| U_{\mu3} \right|^2
$.
Thus,
oscillations of terrestrial neutrinos
are characterized in this scheme by three positive parameters:
$ \Delta m^2_{31} \equiv \Delta m^2 $,
$ \left| U_{e3} \right|^2 $
and
$ \left| U_{\mu3} \right|^2 $.

Let us consider first the results of
reactor and accelerator disappearance experiments
in which
$ \bar\nu_e \to \bar\nu_x $
and
$ \nu_\mu \to \nu_x $
transitions were searched for.
No indications in favor of
neutrino oscillations were found in these experiments.
We will use the exclusion plots
which depict the limits obtained
in the recent Bugey reactor experiment
\cite{BUGEY95}
and in the
CDHS and CCFR accelerator experiments
\cite{CDHS84,CCFR84}.
At fixed values of
$ \Delta m^2 $,
the allowed values of
the amplitudes
$ B_{\nu_{e};\nu_{e}} $
and
$ B_{\nu_{\mu};\nu_{\mu}} $
are constrained by
\begin{equation}
\begin{array}{l} \displaystyle
B_{\nu_{e};\nu_{e}}
\le
B_{\nu_{e};\nu_{e}}^{0}
\;,
\\ \displaystyle
\\ \displaystyle
B_{\nu_{\mu};\nu_{\mu}}
\le
B_{\nu_{\mu};\nu_{\mu}}^{0}
\;.
\end{array}
\label{E141}
\end{equation}
The values of
$ B_{\nu_{e};\nu_{e}}^{0} $
and
$ B_{\nu_{\mu};\nu_{\mu}}^{0} $
can be found from the corresponding exclusion curves.

{}From Eq.(\ref{E110})
the parameters
$ \left| U_{\alpha3} \right|^2 $
(with $\alpha=e,\mu$)
can be expressed in terms of the amplitudes
$ B_{\nu_{\alpha};\nu_{\alpha}} $
as
\begin{equation}
\left| U_{\alpha3} \right|^2
=
{1\over2}
\left(
1
\pm
\sqrt{ 1 - B_{\nu_{\alpha};\nu_{\alpha}} }
\right).
\label{E143}
\end{equation}
Thus,
from the negative results
of reactor and accelerator disappearance experiments
we see that the parameters
$ \left| U_{\alpha3} \right|^2 $
at fixed values of
$ \Delta m^2 $
must satisfy one of the following inequalities:
\begin{equation}
\begin{array}{l} \displaystyle
\left| U_{\alpha3} \right|^2
\ge
{\textstyle{1\over2}}
\left(
1
+
\sqrt{ 1 - B_{\nu_{\alpha};\nu_{\alpha}}^{0} }
\right)
\equiv
a_{\alpha}^{(+)}
\\ \displaystyle
\hskip-2cm
\mbox{or}
\\ \displaystyle
\left| U_{\alpha3} \right|^2
\le
{\textstyle{1\over2}}
\left(
1
-
\sqrt{ 1 - B_{\nu_{\alpha};\nu_{\alpha}}^{0} }
\right)
\equiv
a_{\alpha}^{(-)}
\;.
\end{array}
\label{E144}
\end{equation}
In Table \ref{TAB1}
we present the values of
$ a_{e}^{(\pm)} $
and
$ a_{\mu}^{(\pm)} $
obtained from the negative results of
the Bugey, CDHS and CCFR experiments
for some values of
$ \Delta m^2 $
in the interval
$ 10^{-1} \, \mathrm{eV}^2 \le \Delta m^2 \le 10^{2} \, \mathrm{eV}^2 $,
that covers the range
where positive indications
in favor of
$ \nu_\mu \leftrightarrows \nu_e $
oscillations
were reported by the LSND experiment
\cite{LSND}.
It can be seen from Table \ref{TAB1}
that in the region of
$ \Delta m^2 $
under consideration
the values of
$ a_{e}^{(-)} $
and
$ a_{\mu}^{(-)} $
are small
whereas the values of
$ a_{e}^{(+)} $
and
$ a_{\mu}^{(+)} $
are large
(close to 1).

{}From Eq.(\ref{E144})
it follows that
for any fixed value of
$ \Delta m^2 $
we have four regions
of possible values of the parameters
$ \left| U_{e3} \right|^2 $
and
$ \left| U_{\mu3} \right|^2 $.
We will consider now all these regions.

\def\theenumi{\Roman{enumi}}

\begin{figure}[p]
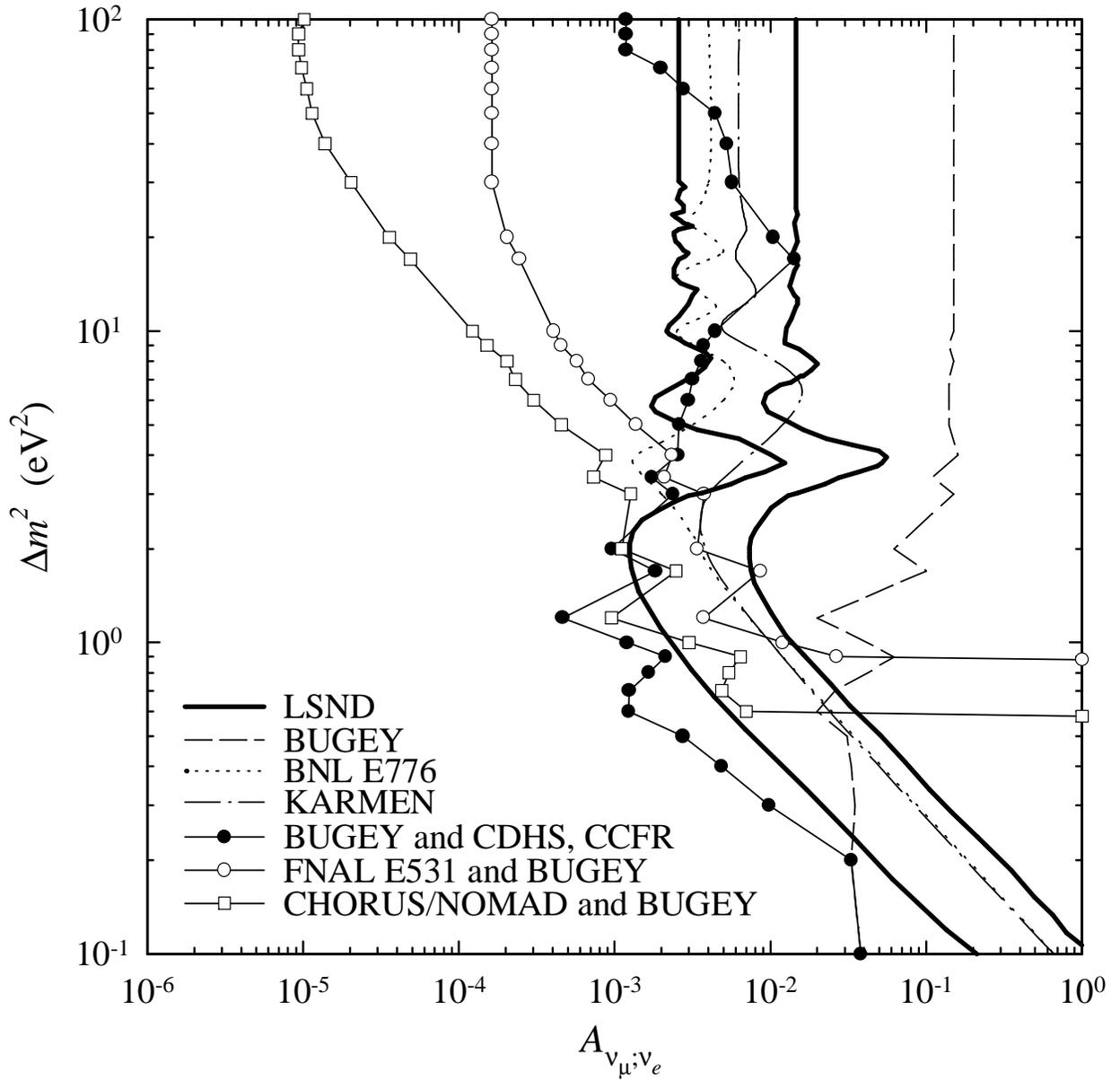

\protect\caption{
The allowed regions in the
$ \Delta m^2 $--$ A_{\nu_{\mu};\nu_{e}} $
plane
obtained from the results of
disappearance reactor and accelerator experiments
and
$ \nu_\mu \leftrightarrows \nu_\tau $
appearance experiments
for small values of
$ \left| U_{e3} \right|^2 $
and
$ \left| U_{\mu3} \right|^2 $.
$ A_{\nu_{\mu};\nu_{e}} $
is the amplitude of $ \nu_\mu \leftrightarrows \nu_e $
oscillations.
The results from
the BNL E776 and KARMEN
$ \nu_\mu \leftrightarrows \nu_e $
appearance experiments
are also drawn.
The region allowed by the LSND experiment
is limited by the two thick solid curves.}
\label{FIG1}
\end{figure}

\begin{enumerate}

\item
\label{REGIONI}
The region of small
$ \left| U_{e3} \right|^2 $
and
$ \left| U_{\mu3} \right|^2 $:
\begin{equation}
\left| U_{e3} \right|^2 \le a_{e}^{(-)}
\qquad \mbox{and} \qquad
\left| U_{\mu3} \right|^2 \le a_{\mu}^{(-)}
\label{E150}
\end{equation}

The region of small values
of the non-diagonal elements
of the mixing matrix
is the most interesting one
from a theoretical point of view for
only in this region the hierarchy
of couplings among lepton generations
can be realized.

To be specific, let us consider, as an example, the value
$ \Delta m^2 = 6 \, \mathrm{eV}^2 $.
The allowed values of the parameters
$ \left| U_{e3} \right|^2 $
and
$ \left| U_{\mu3} \right|^2 $
are given by (see Table \ref{TAB1})
\begin{equation}
\left| U_{e3} \right|^2
\le
3.6 \times 10^{-2}
\null \quad \mbox{and} \quad \null
\left| U_{\mu3} \right|^2
\le
2.0 \times 10^{-2}
\;.
\label{E151}
\end{equation}
{}From these inequalities
we obtain the following upper bound
for the amplitude
$ A_{\nu_{\mu};\nu_{e}} $
of
$ \nu_\mu \leftrightarrows \nu_e $
oscillations:
\begin{equation}
A_{\nu_{\mu};\nu_{e}}
\le
2.9 \times 10^{-3}
\;.
\label{E152}
\end{equation}

Additional restrictions for the allowed values
of the amplitude
$ A_{\nu_{\mu};\nu_{e}} $
can be obtained from the data
of the experiments searching for
$ \nu_\mu \to \nu_\tau $
oscillations.
In the region under consideration,
from Eq.(\ref{E108}) we have
\begin{equation}
\left| U_{\mu3} \right|^2
\lesssim
{\textstyle{1\over4}}
A_{\nu_{\mu};\nu_{\tau}}^{0}
\;,
\label{E206}
\end{equation}
where
$ A_{\nu_{\mu};\nu_{\tau}}^{0} $
is the upper bound
of the amplitude of
$ \nu_\mu \leftrightarrows \nu_\tau $
oscillations.
{}From the exclusion plot of the FNAL E531 experiment
\cite{E531}
which implies,
at
$ \Delta m^2 = 6 \, \mathrm{eV}^2 $,
$ A_{\nu_{\mu};\nu_{\tau}}^{0} = 2.5 \times 10^{-2} $,
we obtain from
Eq.(\ref{E206})
$ \left| U_{\mu3} \right|^2 \lesssim 6 \times 10^{-3} $.
This value,
combined with the bound for
$ \left| U_{e3} \right|^2 $
in Eq.(\ref{E151}),
gives
\begin{equation}
A_{\nu_{\mu};\nu_{e}}
\lesssim
9 \times 10^{-4}
\;.
\label{E153}
\end{equation}
This upper bound is smaller than the bounds
found in the experiments searching for
$ \nu_\mu \to \nu_e $
transitions
and
the values of
$ A_{\nu_{\mu};\nu_{e}} $
reported by the LSND experiment
(see Fig.\ref{FIG1}).
Thus,
$ \nu_\mu \leftrightarrows \nu_e $
oscillations
with an amplitude larger than
$ 9 \times 10^{-4} $
at
$ \Delta m^2 = 6 \, \mathrm{eV}^2 $,
are forbidden if both the parameters
$ \left| U_{e3} \right|^2 $
and
$ \left| U_{\mu3} \right|^2 $
are small.

In Fig.\ref{FIG1}
we have shown the upper bounds for the amplitude
$ A_{\nu_{\mu};\nu_{e}} $
obtained from the exclusion plots
of reactor and accelerator experiments
for the case of small
$ \left| U_{e3} \right|^2 $
and
$ \left| U_{\mu3} \right|^2 $
in the range
$ 10^{-1}\mathrm{eV}^2 \le \Delta m^2 \le 10^{2} \, \mathrm{eV}^2 $.
The curve passing through the filled circles
was obtained from the exclusion plots of the
Bugey \cite{BUGEY95},
CDHS \cite{CDHS84}
and
CCFR \cite{CCFR84}
experiments.
The curve passing through the open circles
was obtained by combining the results of the
Bugey experiment
and the FNAL E531 \cite{E531}
experiment,
which was searching for
$ \nu_\mu \to \nu_\tau $
transitions,
as explained above for the case of
$ \Delta m^2 = 6 \, \mathrm{eV}^2 $.
In Fig.\ref{FIG1}
we have also plotted parts of the exclusion curves
obtained by the
KARMEN
\cite{KARMEN}
(dash-dotted line)
and
BNL E776
\cite{E776}
(dotted line)
experiments
which were searching for
$ \nu_\mu \to \nu_e $
transitions.
Finally,
taking into account that
$
A_{\nu_{\mu};\nu_{e}}
\le
B_{\nu_{e};\nu_{e}}
$
(see Eq.(\ref{E112})),
we also plotted in Fig.\ref{FIG1}
the exclusion curve
for
$ B_{\nu_{e};\nu_{e}} $
found in the Bugey experiment
(dashed line).

It can be seen from Fig.\ref{FIG1}
that in the region of
$ \Delta m^2 $ under consideration
the results of
reactor $\bar\nu_e$ and accelerator $\nu_\mu$
disappearance experiments
combined with those of
$ \nu_\mu \to \nu_\tau $
appearance experiments
provide more severe restrictions
on the allowed values of the amplitude
$ A_{\nu_{\mu};\nu_{e}} $
than the results of direct
$ \nu_\mu \to \nu_e $
appearance experiments.
If any experiment searching for
$ \nu_\mu \leftrightarrows \nu_e $
oscillations
finds,
for
$ \Delta m^2 $
in the range under consideration,
an amplitude
$ A_{\nu_{\mu};\nu_{e}} $
larger than the upper bound
presented in Fig.\ref{FIG1},
it would mean that
the parameters
$ \left| U_{e3} \right|^2 $
and
$ \left| U_{\mu3} \right|^2 $
cannot be both small,
i.e.
there is no natural hierarchy
of generation couplings in the lepton sector.
Such would be the case
if the LSND result is confirmed,
as it is seen from Fig.\ref{FIG1},
where the LSND allowed region
is limited by the two thick solid lines.

The new experiments
CHORUS
\cite{CHORUS}
and NOMAD
\cite{NOMAD}
searching for
$ \nu_\mu \to \nu_\tau $
transitions
are under way at CERN.
In Fig.\ref{FIG1}
we have also plotted the range of
$ A_{\nu_{\mu};\nu_{e}} $
that could be explored
when the projected sensitivity of the
CHORUS and NOMAD
experiments is reached
(taking into account the bounds on
$ \left| U_{e3} \right|^2 $
obtained from the Bugey experiment).

\item
\label{REGIONII}
The region of large
$ \left| U_{e3} \right|^2 $
and
$ \left| U_{\mu3} \right|^2 $:
\begin{equation}
\left| U_{e3} \right|^2 \ge a_{e}^{(+)}
\qquad \mbox{and} \qquad
\left| U_{\mu3} \right|^2 \ge a_{\mu}^{(+)}.
\label{E160}
\end{equation}

The unitarity of the mixing matrix
implies that
$
\left| U_{e3} \right|^2
+
\left| U_{\mu3} \right|^2
\le
1
$.
It can be seen from Table \ref{TAB1}
that
in the range of
$ \Delta m^2 $ under consideration
the inequality
$
a_{e}^{(+)}
+
a_{\mu}^{(+)}
>
1
$ always holds.
Therefore,
the values of the parameters
$ \left| U_{e3} \right|^2 $
and
$ \left| U_{\mu3} \right|^2 $
cannot be both large,
implying that this region is ruled out.

\item
\label{REGIONIII}
The region of large
$ \left| U_{e3} \right|^2 $
and small
$ \left| U_{\mu3} \right|^2 $:
\begin{equation}
\left| U_{e3} \right|^2 \ge a_{e}^{(+)}
\qquad \mbox{and} \qquad
\left| U_{\mu3} \right|^2 \le a_{\mu}^{(-)}
\label{E170}
\end{equation}

If the neutrino masses satisfy the hierarchy relation (\ref{E103}),
the survival probability of the solar neutrinos
is given by
\cite{SOLARTHREEGEN}
\begin{equation}
P_{\nu_e\to\nu_e}
=
\left(
1
-
\left| U_{e3} \right|^2
\right)^2
P_{\nu_e\to\nu_e}^{(1,2)}
+
\left| U_{e3} \right|^4
\;,
\label{E172}
\end{equation}
where
$ P_{\nu_e\to\nu_e}^{(1,2)} $
is the survival probability
due to the mixing between
the first and the second generations.
Using the values of
$ a_{e}^{(+)} $
listed in Table \ref{TAB1},
we find that
$
P_{\nu_e\to\nu_e}
\ge
0.92
$
for all values of the neutrino energy.
This lower bound
is not compatible
with the data of solar neutrino experiments
\cite{SOLAREXP},
including the data of
GALLEX
and
SAGE which have shown less neutrino deficit than
the Homestake and Kamiokande experiments.
Therefore,
the allowed values of the parameters
$ \left| U_{e3} \right|^2 $
and
$ \left| U_{\mu3} \right|^2 $
cannot be in this region.

\item
\label{REGIONIV}
The region of small
$ \left| U_{e3} \right|^2 $
and large
$ \left| U_{\mu3} \right|^2 $:
\begin{equation}
\left| U_{e3} \right|^2 \le a_{e}^{(-)}
\quad \mbox{and} \quad
\left| U_{\mu3} \right|^2 \ge a_{\mu}^{(+)}
\;.
\label{E205}
\end{equation}
The allowed region for the parameters
$ \left| U_{e3} \right|^2 $
and
$ \left| U_{\mu3} \right|^2 $
is considerably narrowed by the bounds obtained
in the experiments searching for
$ \nu_\mu \to \nu_\tau $
transitions.
In fact,
using relation (\ref{E108})
we obtain
\begin{equation}
\left| U_{\mu3} \right|^2
=
{1\over2}
\left(
1
-
\left| U_{e3} \right|^2
\pm
\sqrt{
\left(
1
-
\left| U_{e3} \right|^2
\right)^2
-
A_{\nu_{\mu};\nu_{\tau}}
}
\right)
\label{E181}
\end{equation}
At fixed values of
$ \Delta m^2 $,
from the corresponding exclusion curves
we obtain
\begin{equation}
A_{\nu_{\mu};\nu_{\tau}}
\le
A_{\nu_{\mu};\nu_{\tau}}^{0}
\;.
\label{E182}
\end{equation}
{}From Eqs.(\ref{E181}) and (\ref{E182})
we find
\begin{equation}
\left| U_{\mu3} \right|^2
\ge
{1\over2}
\left(
1
-
\left| U_{e3} \right|^2
+
\sqrt{
\left(
1
-
\left| U_{e3} \right|^2
\right)^2
-
A_{\nu_{\mu};\nu_{\tau}}^{0}
}
\right)
\label{E183}
\end{equation}

Further,
from the results of the experiments
searching for
$ \nu_\mu \leftrightarrows \nu_e $
oscillations,
at any fixed value of
$ \Delta m^2 $
we have
\begin{equation}
\left| U_{e3} \right|^2
\le
{\displaystyle
A_{\nu_{\mu};\nu_{e}}^{0}
\over\displaystyle
4
\left| U_{\mu3} \right|^2
}
\;,
\label{E184}
\end{equation}
where
the values of
$ A_{\nu_{\mu};\nu_{e}}^{0} $
can be obtained from the exclusion curves
of the experiments
searching for
$ \nu_\mu \to \nu_e $
transitions.

It is also useful to notice that
in the region under consideration
the parameters
$ \left| U_{e3} \right|^2 $
and
$ \left| U_{\mu3} \right|^2 $
are related with the amplitudes
$ A_{\nu_{\mu};\nu_{e}} $
and
$ A_{\nu_{\mu};\nu_{\tau}} $
by the following relations:
\begin{eqnarray}
&&
\left| U_{e3} \right|^2
\simeq
{\textstyle{1\over4}}
\,
A_{\nu_{\mu};\nu_{e}}
\label{E185}
\\
&&
\left| U_{\mu3} \right|^2
\simeq
1
-
{\textstyle{1\over4}}
\left(
A_{\nu_{\mu};\nu_{e}}
+
A_{\nu_{\mu};\nu_{\tau}}
\right)
\label{E186}
\end{eqnarray}
These relations
are valid for small
$ \left| U_{e3} \right|^2 $
and
$ 1 - \left| U_{\mu3} \right|^2 $.

Let us consider as an example
$ \Delta m^2 = 6 \, \mathrm{eV}^2 $.
In Fig.\ref{FIG2}
we plotted the allowed region for
the values of the parameters
$ \left| U_{e3} \right|^2 $
and
$ \left| U_{\mu3} \right|^2 $.
The region (\ref{E205})
is limited by the dash-dotted and dash-dot-dotted lines.
The unitarity limit
$
\left| U_{e3} \right|^2
+
\left| U_{\mu3} \right|^2
\le
1
$
is given by the thick solid curve.
The allowed regions are indicated by the arrows.
Using the value
$ A_{\nu_{\mu};\nu_{\tau}}^{0} = 2.5 \times 10^{-2} $
given by the exclusion curve of the
FNAL E531 experiment
\cite{E531},
from
Eq.(\ref{E183})
we obtained the limiting curve
represented as a dotted line in Fig.\ref{FIG2}.
Using the limits
$ A_{\nu_{\mu};\nu_{e}} \le 5.0 \times 10^{-3} $
and
$ A_{\nu_{\mu};\nu_{e}} \le 1.4 \times 10^{-2} $
given by the BNL E776
\cite{E776}
and
KARMEN
\cite{KARMEN}
experiments,
from
Eq.(\ref{E185})
we obtained the two limiting boundaries
that are represented by the solid vertical lines
in Fig.\ref{FIG2}.
The regions A and B
are allowed by all experiments except LSND.
In Fig.\ref{FIG2}
we have also shown the region allowed by the LSND experiment
(between the two thick solid lines).
If we take into account the LSND result,
only the region B is allowed.

If the parameters
$ \left| U_{e3} \right|^2 $
and
$ \left| U_{\mu3} \right|^2 $
satisfy the inequalities (\ref{E205}),
the results of disappearance experiments
and the experiments searching for
$ \nu_\mu \leftrightarrows \nu_\tau $
oscillations
do not provide any constraint on
$ \nu_\mu \leftrightarrows \nu_e $
oscillations.
If,
for example,
$ \nu_\mu \leftrightarrows \nu_\tau $
oscillations
are found by the CHORUS and NOMAD experiments
with an amplitude
$ A_{\nu_{\mu};\nu_{\tau}} $,
it would mean that the parameters
$ \left| U_{e3} \right|^2 $
and
$ \left| U_{\mu3} \right|^2 $
are related by Eq.(\ref{E181})
(with the plus sign),
but the amplitude
$ A_{\nu_{\mu};\nu_{e}} $
of
$ \nu_\mu \leftrightarrows \nu_e $
oscillations
would not be constrained by this result.
On the other hand,
the observation of
$ \nu_\mu \leftrightarrows \nu_e $
oscillations
with an amplitude
$ A_{\nu_{\mu};\nu_{e}} $
would  allow us to determine
$ \left| U_{e3} \right|^2 $
(see Eq.(\ref{E185})).
However,
as it is seen from Eqs.(\ref{E185}) and (\ref{E186})
the amplitude
$ A_{\nu_{\mu};\nu_{\tau}} $
of
$ \nu_\mu \leftrightarrows \nu_\tau $
oscillations
would not be constrained by this result.
The same considerations apply to other values of
$ \Delta m^2 $.

Let us notice that
if a positive signal is found
in the experiments searching for
$ \nu_\mu \leftrightarrows \nu_e $
oscillations
and in the experiments searching for
$ \nu_\mu \leftrightarrows \nu_\tau $
oscillations,
the values of both parameters
$ \left| U_{e3} \right|^2 $
and
$ \left| U_{\mu3} \right|^2 $
would be determined unambiguously.

As is well known,
if massive neutrinos
are Majorana particles,
neutrinoless double beta decay
is allowed.
The matrix elements of this process
are proportional to
$ \displaystyle
\langle m \rangle
=
\sum_{i}
U_{ei}^2
\,
m_{i}
$.
If,
for example,
$ \Delta m^2 = 6 \, \mathrm{eV}^2 $,
from Fig.\ref{FIG2}
we find the upper bound
$ \displaystyle
\left| \langle m \rangle \right|
\lesssim
3 \times 10^{-3} \, \mathrm{eV}
$,
which is much less
than the sensitivity
reachable by future experiments
(see Ref.\cite{DOUBLEBETA}).

\end{enumerate}

Thus,
if the model
with mixing of three generations of neutrinos
and a neutrino mass hierarchy
turns out to be correct,
the observation of a positive signal in
$ \nu_\mu \leftrightarrows \nu_e $
oscillation experiments
with an amplitude larger than
the limit given in Fig.\ref{FIG1},
would imply that
$ \left| U_{\mu3} \right|^2 $
is large
and
$ \left| U_{e3} \right|^2 $
is small.
This result would mean that
the neutrino mixing matrix
is basically different
from the Cabibbo-Kobayashi-Maskawa mixing matrix of quarks.
We have stressed already that
in the case of neutrino mixing,
a  neutrino with a given flavor does not have a definite mass.
The approximate notion
of ``mass'' of a flavor neutrino
can be introduced
only if one of the corresponding elements
of the mixing matrix is large.
It is obvious that in the region \ref{REGIONI},
where
$ \left| U_{e3} \right|^2 $
and
$ \left| U_{\mu3} \right|^2 $
are both small,
$\nu_\tau$
is the neutrino with largest ``mass''.
If the elements of the mixing matrix
are found to be in the region \ref{REGIONIV},
a very unusual situation
in which
$\nu_\mu$
is the neutrino with the largest ``mass''
would be realized.

\section{Conclusions}
\label{CONCLUSIONS}

We have analyzed the results
of reactor and accelerator neutrino oscillation experiments
in the framework of a general model
with mixing of three neutrino generations
and a natural neutrino mass hierarchy
$ m_1 \ll m_2 \ll m_3 $.
Our only assumption is that
$ \Delta m^2_{21} \equiv m_2^2 - m_1^2 $
is in the range suitable for solving the solar neutrino problem.
In this model
the oscillations of terrestrial neutrinos
are described by three parameters:
$ \Delta m^2 \equiv m_3^2 - m_1^2 $,
$m_3$ being the heaviest mass,
and the squared moduli of two elements of the mixing matrix,
$ \left| U_{e3} \right|^2 $
and
$ \left| U_{\mu3} \right|^2 $.
Using only the results of disappearance experiments,
at any fixed value of
$ \Delta m^2 $
in the range
$ 10^{-1} \, \mathrm{eV}^2 \le \Delta m^2 \le 10^{2} \, \mathrm{eV}^2 $,
we found four possible regions
for the values of the parameters
$ \left| U_{e3} \right|^2 $
and
$ \left| U_{\mu3} \right|^2 $.
We have shown that if both parameters are small,
the limit on the amplitude
$ A_{\nu_{\mu};\nu_{e}} $
of
$ \nu_\mu \leftrightarrows \nu_e $
oscillations
that can be obtained
from the results of reactor and accelerator
disappearance experiments
and from the results of the experiments searching for
$ \nu_\mu \leftrightarrows \nu_\tau $
oscillations
is much less than the limit
on
$ A_{\nu_{\mu};\nu_{e}} $
that was obtained from direct
$ \nu_\mu \leftrightarrows \nu_e $
appearance experiments.
{}From our analysis it follows that
if the LSND signal
is confirmed,
it would mean that
the values of the parameters
$ \left| U_{e3} \right|^2 $
and
$ \left| U_{\mu3} \right|^2 $
cannot be both small,
i.e.
there is no natural hierarchy of couplings among generations
in the lepton sector.
The region
with large
$ \left| U_{e3} \right|^2 $
and
$ \left| U_{\mu3} \right|^2 $
and the region with large
$ \left| U_{e3} \right|^2 $
and small
$ \left| U_{\mu3} \right|^2 $
are excluded by the unitarity condition
and by the results of the solar neutrino experiments,
respectively.
The region with small
$ \left| U_{e3} \right|^2 $
and large
$ \left| U_{\mu3} \right|^2 $
can accommodate the LSND signal.
If this solution is realized in nature,
the LSND signal would not influence
$ \nu_\mu \leftrightarrows \nu_\tau $
oscillations
that are being searched for in the CHORUS and NOMAD experiments.

We have not discussed in this paper
the atmospheric neutrino anomaly
\cite{ATMOSPHERIC}.
If future experiments,
including long-baseline experiments
(see, for example, Ref.\cite{LONGBASELINE}),
confirm
the existence of this anomaly
and  the result of
the LSND experiment
is confirmed also,
three different
$ \Delta m^2 $
scales would be required
for the explanation of all these effects
and the data of the solar neutrino experiments.
In this case
it would be necessary to assume the existence
of an additional sterile neutrino state
besides the three active flavor neutrino states.
The discussion of models
with three active and sterile states
is out of the scope of this paper.
Some models of this type
have been recently considered
\cite{STERILE}.

\acknowledgments

It is a pleasure for us
to express our gratitude
to
Milla Baldo Ceolin and
Alberto Marchionni
for very useful discussions.


\widetext

\begin{table}[p]
\protect\caption{Values of the upper and lower bounds
$ a_{\alpha}^{(-)} $
and
$ a_{\alpha}^{(+)} $
($\alpha=e,\mu$)
for the parameter
$ \left| U_{\alpha3} \right|^2 $
found from the results
of reactor and accelerator disappearance experiments
for some values of
$ \Delta m^2 $
in the range
$ 10^{-1} \le \Delta m^2 \le 10^{2} \, \mathrm{eV}^2 $.}
\begin{tabular}{ccccc}
$ \Delta m^2 \, \mathrm{(eV)^2}$
&
$ a_{e}^{(-)} $
&
$ a_{e}^{(+)} $
&
$ a_{\mu}^{(-)} $
&
$ a_{\mu}^{(+)} $
\\
\hline
0.1 & 0.0096 & 0.9904 & --- & ---
\\
0.2 & 0.0083 & 0.9917 & --- & ---
\\
0.3 & 0.0088 & 0.9912 & 0.28 & 0.72
\\
0.4 & 0.0083 & 0.9917 & 0.15 & 0.85
\\
0.5 & 0.0078 & 0.9922 & 0.088 & 0.912
\\
0.6 & 0.0050 & 0.9950 & 0.061 & 0.939
\\
0.7 & 0.0065 & 0.9935 & 0.047 & 0.953
\\
0.8 & 0.011 & 0.989 & 0.039 & 0.961
\\
0.9 & 0.016 & 0.984 & 0.034 & 0.966
\\
1 & 0.011 & 0.989 & 0.028 & 0.972
\\
2 & 0.016 & 0.984 & 0.015 & 0.985
\\
3 & 0.039 & 0.961 & 0.015 & 0.985
\\
4 & 0.042 & 0.958 & 0.015 & 0.985
\\
5 & 0.036 & 0.964 & 0.018 & 0.982
\\
6 & 0.036 & 0.964 & 0.020 & 0.980
\\
7 & 0.036 & 0.964 & 0.022 & 0.978
\\
8 & 0.039 & 0.961 & 0.023 & 0.977
\\
9 & 0.036 & 0.964 & 0.026 & 0.974
\\
10 & 0.039 & 0.961 & 0.028 & 0.972
\\
20 & 0.039 & 0.961 & 0.067 & 0.933
\\
30 & 0.039 & 0.961 & 0.036 & 0.964
\\
40 & 0.039 & 0.961 & 0.034 & 0.966
\\
50 & 0.039 & 0.961 & 0.028 & 0.972
\\
60 & 0.039 & 0.961 & 0.018 & 0.982
\\
70 & 0.039 & 0.961 & 0.013 & 0.987
\\
80 & 0.039 & 0.961 & 0.0076 & 0.9924
\\
90 & 0.039 & 0.961 & 0.0076 & 0.9924
\\
100 & 0.039 & 0.961 & 0.0076 & 0.9924
\\
\end{tabular}
\label{TAB1}
\end{table}

\narrowtext

\begin{figure}[p]
\protect\caption{The allowed region
for the values of the parameters
$ \left| U_{e3} \right|^2 $
and
$ \left| U_{\mu3} \right|^2 $
for
$ \Delta m^2 = 6 \, \mathrm{eV}^2 $
in the region of small
$ \left| U_{e3} \right|^2 $
and large
$ \left| U_{\mu3} \right|^2 $.
The region allowed by the LSND experiment
is limited by the two thick solid lines.}
\label{FIG2}
\end{figure}

\newpage
\setcounter{figure}{0}

\begin{figure}[p]
\begin{center}
\mbox{\epsfig{file=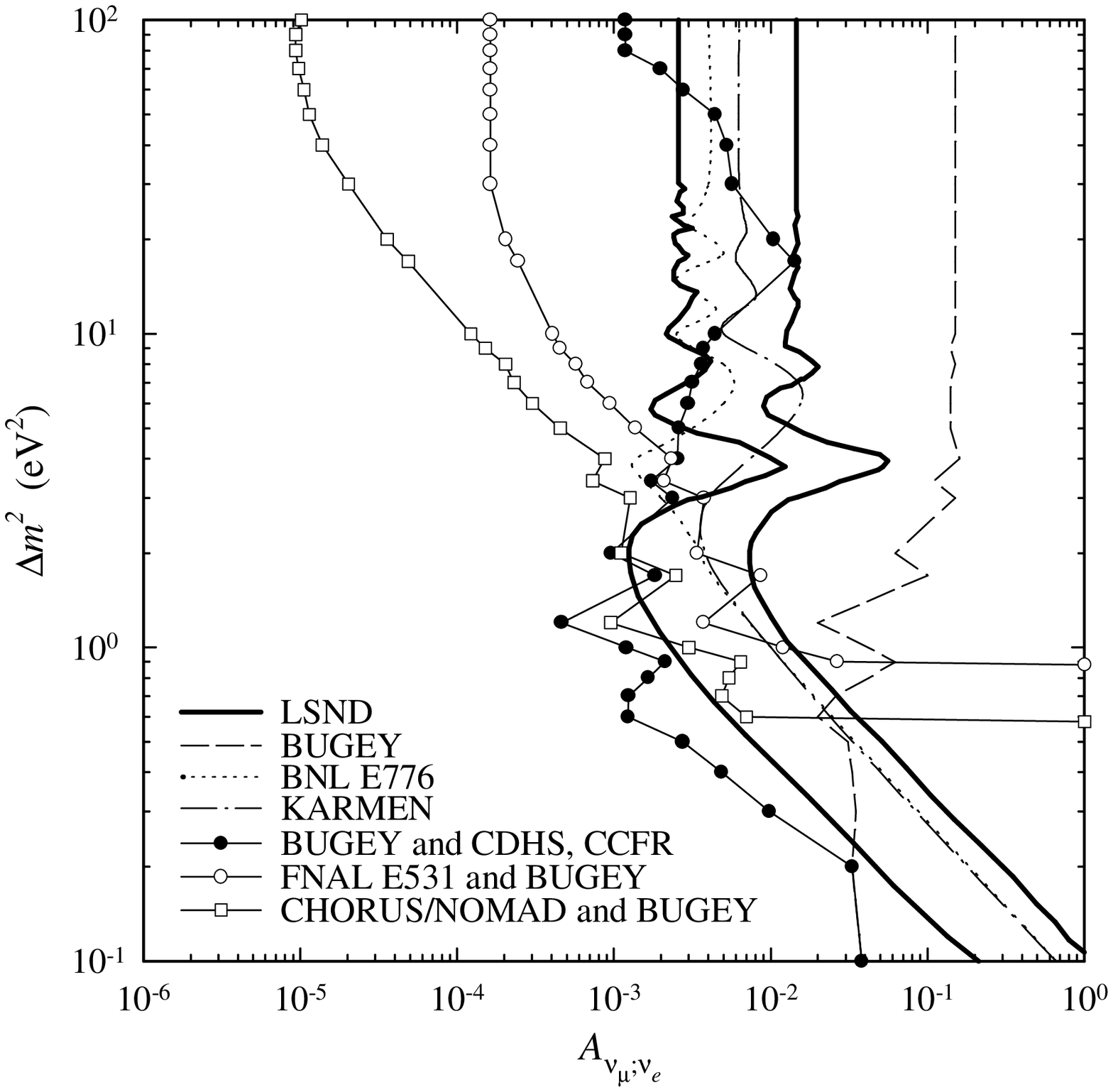,width=\textwidth}}
\end{center}
\protect\caption{
The allowed regions in the
$ \Delta m^2 $--$ A_{\nu_{\mu};\nu_{e}} $
plane
obtained from the results of
disappearance reactor and accelerator experiments
and
$ \nu_\mu \leftrightarrows \nu_\tau $
appearance experiments
for small values of
$ \left| U_{e3} \right|^2 $
and
$ \left| U_{\mu3} \right|^2 $.
$ A_{\nu_{\mu};\nu_{e}} $
is the amplitude of $ \nu_\mu \leftrightarrows \nu_e $
oscillations.
The results from
the BNL E776 and KARMEN
$ \nu_\mu \leftrightarrows \nu_e $
appearance experiments
are also drawn.
The region allowed by the LSND experiment
is limited by the two thick solid curves.}
\end{figure}

\begin{figure}[p]
\begin{center}
\mbox{\epsfig{file=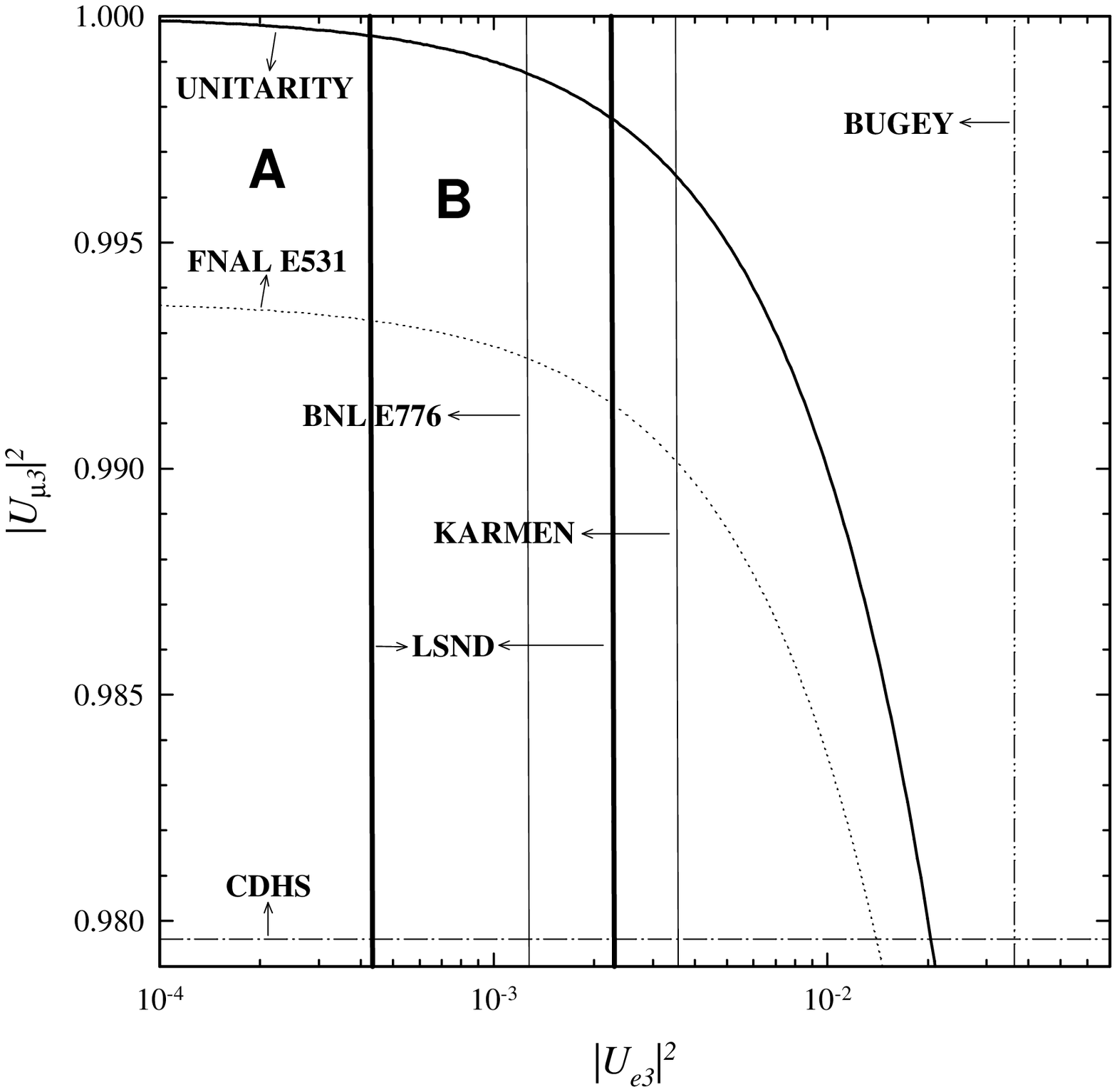,width=\textwidth}}
\end{center}
\protect\caption{The allowed region
for the values of the parameters
$ \left| U_{e3} \right|^2 $
and
$ \left| U_{\mu3} \right|^2 $
for
$ \Delta m^2 = 6 \, \mathrm{eV}^2 $
in the region of small
$ \left| U_{e3} \right|^2 $
and large
$ \left| U_{\mu3} \right|^2 $.
The region allowed by the LSND experiment
is limited by the two thick solid lines.}
\end{figure}

\end{document}